\documentclass[twocolumn,prd,floatfix,nofootinbib]{revtex4}
\usepackage{mathrsfs}
\usepackage{bbm}
\usepackage{amsmath}
\usepackage{graphicx}
\usepackage{dcolumn}
\usepackage{bm}
\usepackage{amssymb}
\usepackage{latexsym}

\begin{document}

\title{Preheating in Bubble Collisions}

\author{Jun Zhang\footnote{Email: junzhang34@Gmail.com}}
\author{Yun-Song Piao\footnote{Email: yspiao@gucas.ac.cn}}

\affiliation{College of Physical Sciences, Graduate School of
Chinese Academy of Sciences, Beijing 100049, China}



\begin{abstract}

In a landscape with metastable minima, the bubbles will inevitably
nucleate. We show that during the bubbles collide, due to the
dramatically oscillating of the field at the collision region, the
energy deposited in the bubble walls can be efficiently released
by the explosive production of the particles. In this sense, the
collision of bubbles is actually highly inelastic. The
cosmological implications of this result are discussed.

\end{abstract}

\maketitle

When the universe is initially set in a metastable minimum of
certain landscape of scalar fields, it will undergo a dS expansion,
bubbles with lower energy minima will inevitably nucleate in this
background \cite{Coleman}. When the radius of bubble is larger than
its critical radius, the bubble will expand outwards, and eventually
collide with other expanding bubbles. In general, it is expected
that during the bubbles collide, the energy deposited in the bubble
walls will be released, e.g.\cite{GW}. This release of energy is
significant, e.g. in old inflation \cite{Guth81}, extended inflation
\cite{LS}, and \cite{Linde90},\cite{AF},\cite{CLL},\cite{LPS}, the
universe is reheated by it.

The bubble collision has been studied earlier in
\cite{Hawking},\cite{Wu}. In general, when the bubbles collide,
the walls of bubbles will pass through each other or be reflected.
The region between outgoing walls remains in high energy
metastable minimum, while other region is not affected. However,
there is a net force, which will compel the walls to rest, and
then back and move towards each other. Thus the collision of walls
will inevitably occur again and again. This oscillation of walls
has be displayed in the numerical simulations for bubble collision
\cite{Hawking},\cite{WW},\cite{KTW},\cite{AJ2}. In general, it is
thought that during the bubble collision the energy deposited in
the walls will be released by the direct decaying of scalar wave
into other particles, e.g.\cite{WW},\cite{KR}, or the
gravitational radiation,
e.g.\cite{KTW},\cite{KT},\cite{CDS},\cite{C}. However, this
release of the energy might be more dramatic than expected. In
this paper, we show that due to the oscillation of the background
field at the collision region, the energy can be efficiently
released by the explosive production of the particles.

We begin with a brief review of the numerically simulation of the
collision of bubbles nucleating in a given potential in Fig.1, in
which the high energy metastable minimum is $\phi_F$ and the low
energy minimum is $\phi_T$. We only care the numerical results of
the evolution of field at the collision region during the bubble
collision. Thus
the details of the equations of the nucleation and evolution of
bubble are neglected, see e.g.\cite{CKL},\cite{AJS}.

\begin{figure}
\includegraphics[width=7.0cm]{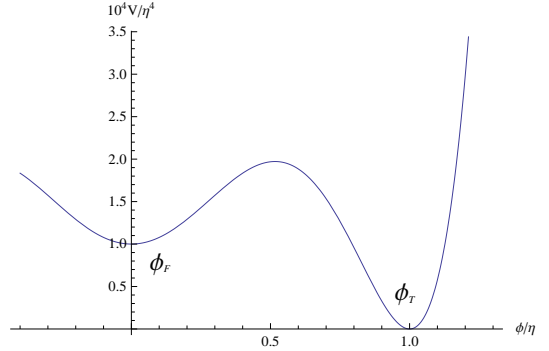}
\caption{\label{fig:v1} The potential with two minima, which will
be used in the numerical simulations of the bubble collision in
Figs.2 and 4, and the production of $\chi$-particle in collision
region in Fig.3. $\eta \equiv |\phi_T-\phi_F|$ is a normalized
parameter with mass dimension. The parameter we used is
$\eta=0.01$ for whole paper. }
\end{figure}

The field $\phi$ is initially in $\phi_F$. Thus the universe is
inflating. Then the bubbles with $\phi_T$ will be expected to
nucleate. The radius of bubble is determined by the instanton
equation of $\phi$. We have numerically solved this instanton
equation and will use the data obtained as the initial state of
bubbles which will collide. Their initial distance is defined as
$D_0$. Here we choose $D_0 \sim {375\over \eta}< {1\over H_F}$,
where $H_F$ is the Hubble rate of the false vaccum which can be
estimated by $H_F \simeq \sqrt{V(\phi_F)}/M_P$ and in this paper
$H_F \simeq \eta^2/100$. $M_P=1$ is set for whole paper and $\eta$
is the normalized parameter with mass dimension, see Fig.1. The
nucleation radius of the bubbles is $R_0={3\sigma/V_F}$, which can
be given by
\begin{eqnarray}\label{R0}
R_0=\frac{3\int^{\phi_T}_{\phi_F}\sqrt{2V\left(\phi\right)}d\phi}{V_F}
\sim 0.3\frac{3\sqrt{2}\eta}{\sqrt{V_F}},
\end{eqnarray}
where $\sigma$ is the surface tension of bubble wall, which means
that $H_FR_0 \sim 3\sqrt{2}\eta \ll 1$, since $\eta\ll 1$ and in
the simulations $\eta=0.01 $ is actually used.

\begin{figure}[t]
\includegraphics[width=8.0cm]{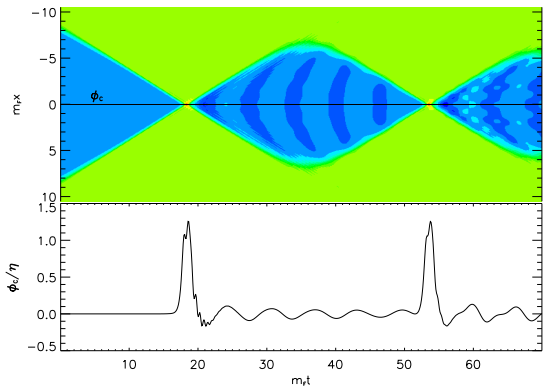}
\caption{\label{fig:phi} The numerical simulation of bubbles
collision. The color panel describes the evolution and collision of
bubbles after their nucleations, where the green region denotes the
region in $\phi_T$ while the blue region denotes that in $\phi_F$,
and the light blue region between them denotes the bubble wall. We
can see that during the bubbles collision the bubble walls will be
oscillating around the center of bubbles collision, which induces
that at the collision region the field $\phi$ initially in $\phi_F$
overshoots to $\phi_T$ and then backs to $\phi_F$, and oscillates
around $\phi_F$. The lower panel is that of $\phi_C$ in color panel.
Here $M_F\sim 0.04\eta$. }
\end{figure}

In the simulation of the evolution and collision of bubbles, we will
neglect the gravitational effect, which is not important since
$V(\phi)\ll 1$, the nucleation radius of bubbles $R_0H_F\ll 1$,
their initial separation $D_0H_F\ll 1$ and the time per oscillation,
which can be found in Fig.\ref{fig:phi}, $50/M_F \simeq 125/\eta \ll
1/H_F$ . We solve the evolving equation $\square\phi=\partial_\phi
V$ of field with the flat space metric in 3+1D by using a modified
version of \textsc{LatticeEasy} \cite{Felder} and 2048 lattices, in
which the initial data of bubbles is given by the instanton
equation, and at $t=0$, $x=\pm 0.5{D_0}$ is considered, since the
collided bubbles can be boosted into a particular frame in which
they are nucleated at the same time. The program is ran with double
precision. The time step is small enough to guarantee the Courant
stability condition. In the meantime, we checked the results by
using higher precision, and found that the results are same. The
numerical result is plotted in Fig.\ref{fig:phi}, in which the width
in space is actually far larger than that in time, and the most of
regions irrelevant with the bubbles collision is cut out to make the
colliding region in figure clearer.

The color panel in Fig.\ref{fig:phi} is that of the evolution and
collision of bubbles in position space. We define the value of
$\phi$ field in a small region around the collision center as
$\phi_C$. The lower panel of Fig.\ref{fig:phi} is the evolution of
$\phi_C$, which can be explained as follows. $\phi_C$ before
collision rests with $\phi_C=\phi_F$. However, when the collide
occurs, the gradient energy of this region will become large, which
will induce $\phi_C$ get cross the potential barrier, overshoot
$\phi_T$. Then it backs to $\phi_F$ and oscillates. This behavior is
repeated during the following collisions of bubble walls. However,
when there is the third lower energy minimum $\phi_3$, after
overshooting $\phi_T$, the field at the collision region might not
back to $\phi_F$, and straightly run into $\phi_3$. Thus a new
bubbles will be generated \cite{Easther}.

We introduce a coupling of the background field $\phi$ with $\chi$
as follows
\begin{eqnarray}\label{cp2}
\mathcal{L}_{int}=-\frac{1}{2}g^2(\phi-\phi_{*})^2\chi^2,
\end{eqnarray}
where $g$ is the coupling constant, and $\phi_{*}$ is a parameter
between $0$ and $1$, see the renormalization in Fig.\ref{fig:v1}.

We will neglect the expansion of space. However, the result is not
altered qualitatively by the inclusion of expansion. The evolution
of $\phi$ at the collision region is given by $\phi_C$ in Fig.2.
This coupling means that when $|\phi_C-\phi_{*}| \sim 0$, the
adiabatic condition becomes violated, the parametric resonance will
inevitably occur, which will result in the explosive production of
$\chi$-particles at corresponding region, similar to the preheating
after inflation \cite{KLS2},\cite{TB}, which has been intensively
studied, e.g.\cite{BTW},\cite{ABCM}.

The initial width $d_0$ of bubble wall can be estimated as
$d_0\sim\frac{3}{2}{\eta\over \sqrt{V_{bar}}}$ \cite{Hawking}, in
which $V_{bar}$ is the effective height of the potential barrier.
This width will become $d=d_0/\gamma$ at the time of bubbles
collision, in which $\gamma$ can be given by $\gamma= {D_0\over
2R_0}+1$, where it is noticed that $D_0$ is the distance between the
bubble walls at the time of their nucleations. When the parameters
in Fig.1 are considered, $d_0\sim 132/\eta$ and $\gamma\sim 2.5$ are
obtained. Thus $d\simeq 53/\eta$. Then, in the linear order the
velocity that $\phi_C$ passes through $\phi_*$ can be estimated as
$\dot{\phi}_*\simeq \Phi_C/d$. Thus in a very short interval $\Delta
t \sim 1/\sqrt{
g{\dot\phi}_*}\simeq\sqrt{\frac{\gamma}{gd_0\Phi_C}}~d$, the
adiabatic condition is broken. Thus the characteristic momentum is
given by
\begin{eqnarray}
k_*\sim (\Delta t)^{-1} \simeq \sqrt{g{\dot\phi}_*}\simeq
\sqrt{g\Phi_C\over d},
\end{eqnarray}
which means $\Delta t \sim 0.35d$ and $k_*\sim 3/d$. This is
consistent with general assumption that the energy of particles
produced by the bubbles collision is about $\sim 1/d$,
e.g.\cite{KR}.

During the bubbles collide, the collision of the bubble walls is
periodical, which induces $|\phi_C-\phi_{*}| \sim 0$ twice at each
time of the collisions of bubble walls, since $\phi_C$ overshoots to
$\phi_T$ and then backs to $\phi_F$. During $\Delta t$ in each time,
$\phi_C(t)$ can be regarded as $\phi_C(t) \simeq {\dot{\phi}}_*
\left(t-t_*\right)$. Thus the motion equation of mode ${X}_k$ of
$\chi$ during $\Delta t$ is given by
\begin{eqnarray}
\frac{d^2 {X}_k}{dt^2}+\left(k^2+{g^2\Phi_C^2\over
d^2}(t-t_*)^2\right){X}_k=0,
\end{eqnarray}
which gives the evolution of ${X}_k$, and thus $n_k$ after each
collision of bubble walls. Thus the occupation number $n_{\chi_k}$
of $X_k$ mode after the $N$th collision of bubble walls, in large
$n_k^i$ limit, is given by $n_{\chi_k}\sim
e^{2\pi\sum\limits_{i=1}^{2N}\mu_k^i}$, where $i$ labels the times
that $\phi_C$ passes through $\phi_*$ and $\mu_k^i$ is the function
of $k/k_*=k/\sqrt{g\Phi_C/d}$ given in \cite{KLS2}. The number
density of $\chi$-particles after the $N$th collision can obtained
by integrating $n_{\chi_k}$ for $k$, which is
\begin{eqnarray}\label{sum}
n_\chi=\frac{1}{8\pi^3}\int d^3k~n_{\chi_k}\sim
\frac{1}{64\pi^3}\left(\frac{g\Phi_C}{d}\right)^{3/2}\frac{e^{2\pi\sum\limits_{i=1}^{2N}\mu_k^i}}{\sqrt{f}},
\end{eqnarray}
which has been evaluated by the steepest descent method, where
$f=\sum\limits_{i=1}^{2N}\mu^i$, $\mu^i$ is the maximum of
$\mu_{k_m}^i(k)$, which can be estimated in $k_m \sim
\sqrt{\frac{g\Phi_C}{4d}}$.

\begin{figure}
\includegraphics[width=8cm]{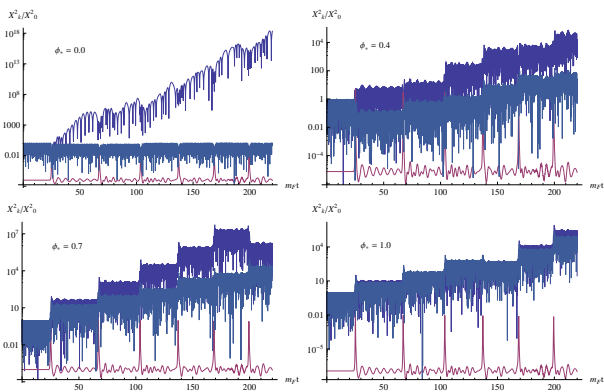}
\caption{\label{fig:chi-s} The numerical simulation of the
evolution of $X_k$ at the collision region for the coupling
(\ref{cp2}). $\eta=0.01$, $g^2=0.5$, and the modes shown are $k=0$
and $k=0.1\eta$, respectively. We also show the evolution of
$\phi_C$ in red lines.}
\end{figure}

We calculate the evolution of ${X}_k$ numerically in
Fig.\ref{fig:chi-s} for different values of $\phi_*$ by using the
data of the bubbles collision in Fig.\ref{fig:chi-s}. The red
lines in this figure denote the evolution of $\phi_C$ during the
bubbles collision. When the bubble walls collide, the red line has
a sharp peak which shows the rapid change of $\phi_C$ overshooting
to $\phi_T$ then backing to $\phi_F$ at the time of each
collision. Here the backreaction from particle production is not
taken into account. However, since during the collisions a little
of energy of the walls will transfer into the energy of
oscillation of the false vacuum, the speed of the walls decrease
after each collision and this will made the peaks of $\Phi_C$ drop
slightly.

In general, the explosive production of particles induced by the
parameter resonance will occur for most $\phi_*$ between 0 and 1.
However, the resonant behavior are different for different
$\phi_*$. The intensity of the parametric resonance is depended on
the speed of $\phi_C$ passes through $\phi_*$. The larger the
speed is, the more likely the broad resonance is to take place. We
can introduce a critical point $\phi_{cri}$ which $\simeq 0.1$ in
our simulation. When $\phi_*\lesssim \phi_{cri}$, the speed of
$\phi_C$ is too small for the broad resonance to take place, and
only narrow resonance occurs, i.e.$X_k$ only increase
exponentially in a band of $k$ with narrow width, see in left
upper panel of Fig.3, in which $X_k$ increases for $k=0$ but is
not changed for $k=0.1\eta$, which is similar to the case in
\cite{TM}. This case is not efficient for the release of energy.
When $\phi_*>\phi_{cri}$, as given in other panels of Fig.3, the
speed of $\phi_C$ become larger, the resonance will be broad, i.e.
it occurs for any values of $k$. It can be found that for the
broad resonance ${X}_k$ is only amplified at the time of each
collision. In this case, the release of energy can be quite
efficient \cite{KLS2}.

The energy density of $\chi$-particles produced after the $N$th
collision is $\rho_{\chi}=n_{\chi}M_{\chi}$, where $M_{\chi}\sim
k_*\simeq \sqrt{g\Phi_C/d}$ can be considered. When the energy
deposited in the bubble walls is completely released, $\rho_\chi$
should be approximately equal to the energy of the corresponding
bubble walls in the collision region. Thus this requires
\begin{eqnarray}\label{eq}
n_{\chi}\sqrt{g\Phi_C\over d}\sim \gamma{\sigma\over l},
\end{eqnarray}
where $R_0$ and the surface tension $\sigma$ are given in
Eq.(\ref{R0}), $\gamma$ has been mentioned, $l$ describes the
effective width of collision region, which is generally $l \simeq
2d$. Thus substituting Eq.(\ref{sum}) into Eq.(\ref{eq}), $N$ can
be estimated as
\begin{eqnarray}\label{N}
 N\sim {1\over 4\pi\mu}\ln{\left({64\pi^3 \sqrt{V_F}\gamma\eta\over l{g^2\Phi_C^2}}\right)},
\end{eqnarray}
where all $\mu^i\simeq \mu$ are adopted. Eq.(\ref{N}) means the
inelastic degree of the bubbles collision is dependent on the
coupling $g$ and the parameters of bubbles at the time of collision,
which is expected. The larger $g$ is, the larger the inelastic
degree is. When the potential given in Fig.1 is introduced,
for $g^2\simeq 0.5$ used in Fig.3, at least $N\simeq 6$ is
required for the complete release of the energy of bubble walls.

It can be significantly noticed that $N$ given by (\ref{N}) is
insensitive to the concrete value of $\phi_*$. The reason is that
the release of energy is proportional to the velocity of $\phi_C$
passing through $\phi_*$, which is about $\sqrt{\Phi_C/d}$
irrelative with $\phi_*$. The simulation of the evolution of field
value at the collision region, in which the loss of walls energy
is considered, is interesting to grasp the physics of the
collision of bubbles. We used the same initial data and parameters
with that in Fig.2, and $\phi_*=0.4$ without loosing
generalization, to perform it in Fig.4. This performance is
explained as follows.

Theoretically, when $\phi$ passes through $\phi_*$, the preheating
leaded by the coupling at $\phi_*$ will make the kinetic energy of
$\phi$ decreasing, and the loss of energy is proportion to the
velocity of $\phi$ at this time. To simulate this effect, we
multiplys a factor $1-\beta^2<1 $ on ${\dot\phi}^2$ in the
corresponding program when $\phi$ pass through $\phi_*$ each time.
$\beta$ can be calculated by letting the loss of kinetic energy of
$\phi$ around $\phi_*$ equal to the energy of particles produced at
this time, which is given by $\beta\simeq {\sqrt{2n^i_{\chi_k}
M_\chi}/ {\dot\phi_C}}$, where $n^i_{\chi_k}$ denotes the number
density of $\chi$-particles produced at each time when $\phi_C\simeq
\phi_*$. In actually simulation, this process is carried out by the
cumulation of the time steps around $\phi_*$. We can see that after
about $N=4$, the blue region denoting $\phi_F$ disappears, $\phi_C$
at the collision region will oscillate around $\phi_T$. The reason
is that with the gradual release of the energy in the bubble walls,
the oscillation of bubble walls will have smaller and smaller
amplitude, which will lead that during the $N$th collision, instead
of backing to $\phi_F$ during previous collision of walls, after
getting cross the potential barrier, $\phi_C$ will oscillate around
$\phi_T$.

In above estimate, the backreaction of $\chi$-particles produced
to the evolution of $\phi$ at the collision region has been
neglected. However, with the increase of $\chi$-particles, its
backreaction will be enhanced gradually, which will inevitably
shut off the resonance at certain time or after some collisions of
bubble walls. The correction induced by the $\chi$-particles to
the potential at the point of $\phi_C\simeq \phi_*$ is $\Delta
M_*^2\simeq g^2<\chi^2>$ \cite{KLS2}, in which the expectation
value of $\chi^2$ is given by $<\chi^2>\simeq {n_{\chi}/
g(\phi_C-\phi_*)}\sim {n_{\chi}/ g\Phi_C}$. The condition that the
backreaction becomes important is $\Delta M^2_*\simeq
V_*^{\prime\prime}$, which gives $n_{\chi}\sim
V_*^{\prime\prime}\Phi_C/g$. Thus substituting Eq.(\ref{sum}), we
can obtain
\begin{eqnarray}\label{reaction}
N_{backreaction}\simeq {1\over 4\pi\mu}\ln{\left({64\pi^3
V_*^{\prime\prime}\sqrt{d^3} \over \sqrt{g^5\Phi_C}}\right)},
\end{eqnarray}
which is $N_{backreaction}\simeq 4$ for the parameters in Fig.1
and $g^2=0.5$. Thus the backreaction is important only after
$N=4$. As displayed in Fig.4, after $N=4$, the metastable $\phi_F$
region has basically disappears, $\phi_C$ at the collision region
will oscillate around $\phi_T$. Thus the simulation in Fig.4
without the backreaction is robust.

\begin{figure}[t]
\includegraphics[width=8cm]{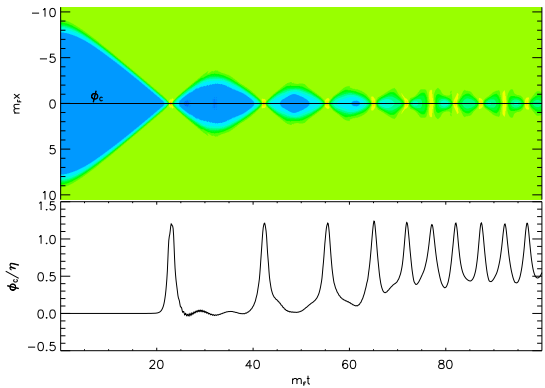}
\caption{\label{fig:chi}The numerical simulation of bubble
collisions, in which the release of walls energy induced by the
production of particles at the time of each collision of bubble
walls is included, $\phi_*=0.4$. The color panel describes the
oscillation of bubble walls during the bubble collision. What the
green region, the blue region and the light blue region denote are
same as that in Fig.2. The lower panel is that of $\phi_C$ in
color panel.}
\end{figure}

When $\phi_C$ oscillates around $\phi_T$, the residual energy is
only that of field oscillation, which will be diluted gradually by
the expansion of space. However, if $\phi_*\simeq 1$, the narrow
resonance around $\phi_T$ will occur during a following short
period, and the residual oscillating energy will continue to be
released into the particles. In a word, eventually $\phi_C$, i.e.the
collision region, will stay at $\phi_T$. The collision of bubbles
ends.

In conclusion, we have proposed a new possibility of the release of
energy deposited in the bubble walls during bubble collisions, not
noticed before. When the bubbles collide, the conventionally
viewpoint is that the energy in the walls is released by the scalar
or gravitational radiation. However, this is not quite efficient. We
show that the energy can be efficiently released by the explosive
production of the particles, due to the parameter resonance. In this
case, the energy in the bubble walls can be drained rapidly. In the
example given, the time that the energy is completely released is
$60/M_F \ll 1/H_F$, and the collision times $N = 4$ is smaller than
$N > 13$ when the preheating is not taken into account which is
usual expected.

This result has interesting applications in extended inflation and
other relevant inflation models, e.g.\cite{AF}. It not only helps
to obtain the enough reheating temperature in these models, but
enrich the phenomenological studies on the reheating of these
models. In principle, the preheating in these models could be
nearly similar to that in slow roll inflation models. Here,
however, the results are dependent on the parameters of bubbles at
the time of collision, which are mainly determined by the
structure of potential landscape. Thus the preheating in these
models could have different predictions from those in slow roll
inflation models.

In this paper, we only consider a simple model. In general,
dependent on the parameters of bubbles at the time of collision,
the oscillation of field at the collision region can be different.
However, as long as the coupling of background field $\phi$ to a
light scalar field $\chi$ is considered, the explosive production
of the particles at the region of bubbles collision will be
general. In a landscape with multiple dimensions, such couplings
can be actually expected, see \cite{Tye:2009ff,Tye:2009rb} for the
studies of cosmological models in such a landscape. This means
that the collision of bubbles is generally highly inelastic. In
principle, the classical $\chi$-wave might be also important for
such a coupling. We left the detailed studies and the discussion
on its correlation with the $\chi$-particles produced, and the
production of particles induced by the collision of bubbles in
different potential landscapes in coming works.

The production of particles makes the bubble collision highly
inelastic. However, it seems not remarkable impact on classical
transition of bubbles \cite{Easther}, since in general the field
excursion is nearly unchanged during the first oscillation.

In principle, it might be possible that for a given potential and
the coupling $g$, the energy can be released completely after a
single collision of bubble walls. The condition that this occurs
is obvious. When the residual energy of the wall after one
collision is smaller than the potential energy between the barrier
and the true vacuum, the field in the collision region do not have
enough energy to cross the barrier between the false vacuum and
the true vacuum, therefore it will stay at the true vacuum and
oscillate around the minimum. In this sense, the collision of
bubbles will be extremely inelastic, and the energy deposited in
the bubble wall can be rapidly drained. Here, the effect of the
coupling on the moving of the bubble walls before the bubble
collision is neglected, since it only slows down the acceleration
of the bubble walls.

We might live inside a bubble universe in eternally inflating
background. Recently, the observable signals of the collision of
bubble universes have been discussed
\cite{CKL},\cite{AJS},\cite{CKL2},\cite{AJ}. The resonant
production of particles during the bubbles collision might bring
some distinct observable signals or impacts on the CMB, which will
be explored in the future.

\textbf{Acknowledgments} We thank Y.F. Cai, Y. Liu for
discussions. This work is supported in part by NSFC under Grant
No:10775180, in part by the Scientific Research Fund of
GUCAS(NO:055101BM03), in part by National Basic Research Program
of China, No:2010CB832805

\end{document}